# A STUDY ON DIGITAL VIDEO BROADCASTING TO A HANDHELD DEVICE (DVB-H), OPERATING IN UHF BAND

Farhat Masood
National University of Sciences and Technology, Pakistan
farhatmasood@hotmail.com

**ABSTRACT**
In this paper, we will understand that the development of the Digital Video Broadcasting to a Handheld (DVB-H) standard makes it possible to deliver live broadcast television to a mobile handheld device. Building upon the strengths of the Digital Video Broadcasting - Terrestrial (DVB-T) standard in use in millions of homes, DVB-H recognizes the trend towards the personal consumption of media.

1. **Broadcasting to Handheld Devices**
   a. Television to a handheld receiver. The concept of providing television-like services on a handheld device has generated much enthusiasm. Already, mobile telecom operators are providing video streaming services using their UMTS networks, or third-generation networks. However, the main alternatives to providing live television services on a handheld device currently available are DMB, ISDB-T, Media-FLO and now DVB-H.
   b. DVB approach to handheld television: DVB-H. Building upon the portable and mobile capabilities of DVB-T, the DVB Project developed the Digital Video Broadcasting on Handhelds (DVB-H) standard for the delivery of audio and video content to mobile handheld devices. DVB-H overcomes two key limitations of the DVBT standard when used for handheld devices - it lowers battery power consumption and improves robustness in the very difficult reception environments of indoor and outdoor portable use in devices with built-in antennas. DVB-H can be used alongside mobile telephone technology and thus benefit from access to a mobile telecom network and a broadcast network.
2. **DVB-H value chain[1]**
   a. Viewers. DVB-H enables viewers to watch television programs on a handheld device. Such handheld televisions are likely to be considered personal items as the act of viewing increasingly becomes an individual, rather than a social, activity. Services can be accessed when viewers are "on the move" – in public transportation, waiting for an appointment or while at work. Hence, handheld viewing will extend the hours of television watching to parts of the day when viewers are not at home.
   b. Broadcasters. With their experience in creating and aggregating content, broadcasters have a privileged role in delivering content for television services to a handheld device. However, broadcasters will need to define the level of their involvement in the DVB-H service offering.
   c. Mobile telecom operators. Mobile telecom operators have access to a large customer database and a sophisticated payment system which can be used for customer billing. Mobile operators have already installed a dense network of cellular transmitter sites which may be helpful to use for the roll-out of DVB-H services.
   d. Broadcast Network Operators. Broadcast network operators have been among the key drivers in the development of the DVB-H standard as they have supported access to network infrastructure. Many of the broadcast networks have been built to provide portable indoor coverage of DVB-T services, the same type of coverage required to support DVB-H services. In addition, the broadcast network operator is well suited to serve as the intermediary between the various service providers.
   e. Manufacturers. Manufacturers of consumer devices and professional system components have actively supported the launch of DVB-H services. The DVB-H prototype chips currently measured have already reached a satisfactory performance level and are expected to be mass produced.
   f. Enablers. Various groups have been working to promote DVB-H and other mobile technologies based upon the DVB-T standard.
3. **Implications for broadcasters[2]**
   a. Screen-size. Handheld screen sizes will resemble those of mobile phones and personal digital assistants (PDAs). They are likely to vary from approximately 5 cm to 12 cm diagonal with a very sharp pixel resolution. Early DVB-H tests of normal TV recoded to around 200kbit/s for display at CIF resolution (352 x 288) offered an enjoyable viewing experience and even permitted the reading of normally sized captions.
b. Cyclical presentations. As a low-cost alternative, broadcasters can re-purpose material for a repeated cyclical presentation. Appropriate material can be edited to the desired length and suitably-sized title and closing elements added. The material is then loaded into a carousel for repeated transmission. Some broadcasters have already begun building experience in the use of cyclical carousel presentation of material over enhanced broadcasting channels.
c. Interactivity. DVB-H television services can benefit from an interaction channel. Although interactivity has not yet been fully exploited for traditional television market, it may be facilitated through the adoption of DVB-H services since a personal device may be more conducive to such activity rather than a shared television set. Interactivity can be used to allow viewer voting, much like the phone call and SMS voting, or even allow viewers to participate in game shows.
d. Current content on UMTS telecom networks. Although it is difficult to compare the service offering available from the UMTS networks while at the early stages of their roll-out, it is useful to understand some of the content currently available.

4. **DVB-H Business Models**
   a. **Model I.** In this model, broadcasters manage the end-relationship with the consumer. The broadcaster receives payments for the use of the service, from consumers from the license fee, or subscription, or through payments made via the telecoms network operator. A variation on this could be broadcast funding from advertising revenue. As this is not an integrated service proposal, consumers may need to pay more than one service provider to obtain the different services. Fully interactive services are a possibility and a separate billing procedure will be necessary for consumers to pay mobile telecom operators for the use of such services.

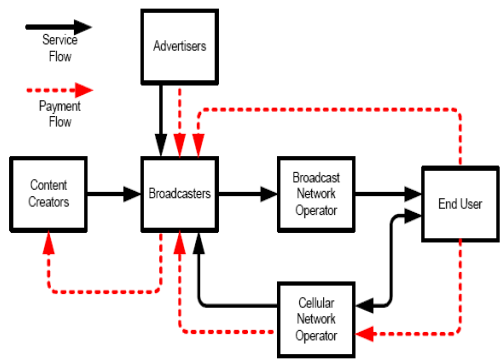

**Figure 1 Broadcaster led approach with Mobile Telecom Operator**

The involvement of the mobile telecom operator may be limited except for linked telecom services. Given the expected initial high cost of DVB-H receivers, market penetration may below if no receiver subsidies are offered

b. **Model – II**. In this model, mobile telecom operators manage the end-relationship with consumers and are responsible for service provisions, marketing and customer care. In addition, mobile telecom operators will need to purchase spectrum and content from broadcasters and other content providers. Consumers have access to an integrated service proposition which means that a complete package will be offered by one service provider.

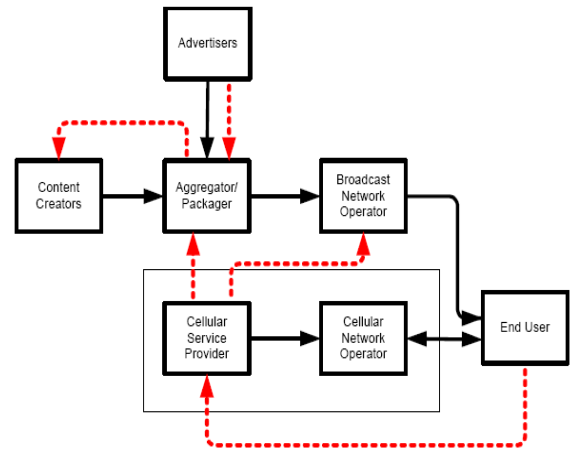

**Figure 2 Mobile telecom operator-led approach with broadcaster**

As a variation, mobile telecom operators could directly handle advertisements. While mobile telecom operators would be responsible for general marketing, it could be possible for broadcasters to market individual television programs. For programs that generate revenue, for example using tele-voting, broadcasters would be responsible for marketing the program while the mobile telecom operator would be responsible for the billing. Revenue would be shared.

c. **Model – III**. In this model, the mobile telecom operator manages the end-relationship with consumers and is responsible for service provisions, marketing and customer care. A dedicated DVB-H service provider acts as a facilitator for mobile operators in the aggregation of content and the use of the spectrum. Consumers have access to an integrated service proposition which means that a complete package will be offered by one service provider.

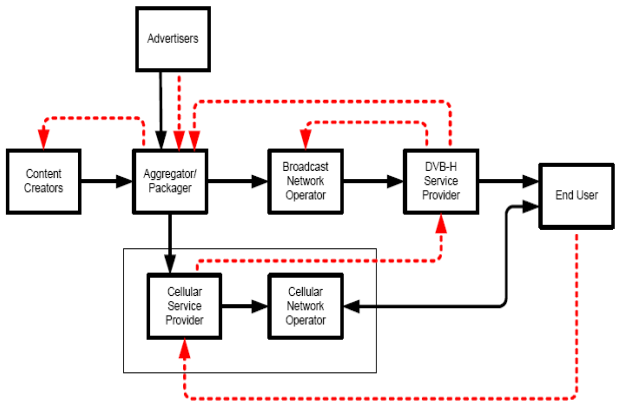

**Figure 3 Independent DVB-H service provider approach**

Variations in this model would include the handling of advertisements directly by the DVBH service provider.

d. **Model – IV**. In this model, the mobile telecom operator is responsible for all aspects of the value chain, from the content creator to the consumer. Broadcasters, or broadcast network operators, provide simply the DVB-H transport capacity. Consumers have access to an integrated service proposition which means that a complete package will be offered by the one service provider. Such a model gives telecom operators a dominant role, with very little involvement from the broadcasting side.

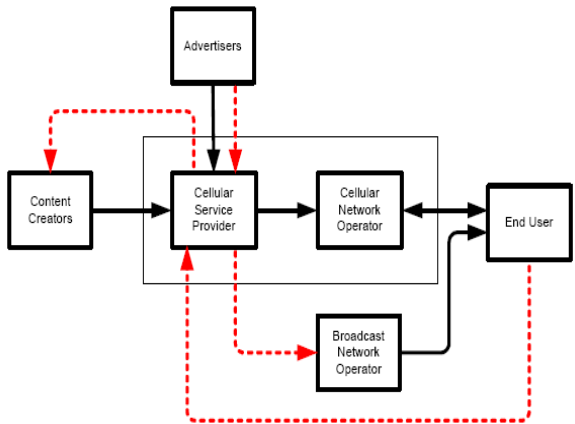

**Figure 4 Mobile telecom operator-led approach**

5. **Technical aspects of DVB-H[3]**
   a. Principles of the DVB-H system. Both DVB-H and DVB-T use the same physical layer and DVB-H can be backwards compatible with DVB-T. Like DVB-T, DVB-H can carry the same MPEG-2 transport stream and use the same transmitter and OFDM modulators for its signal. In addition, DVB-H broadcasts sound, picture and other data using Internet Protocol (IP).
   b. Time-Slicing. To improve the operating time, DVB-H uses time-slicing. Video and audio data (1-2 Mbits), generally representing between 1-5 seconds of the content arrives in the single burst. In addition time-sliced and non time-sliced services can be placed in the same multiplex.
   c. MPE-FEC(Multi-Protocol Encapsulation / Forward Error Correction). Because handheld devices have small antennas that require reception from many different locations, they necessitate a robust transmission system with solid error protection. DVB-H offers improved transmission robustness through the use of an additional level of forward error correction (FEC) at the Multi Protocol Encapsulation (MPE) layer. The use of MPE-FEC is optional.
   d. IPDC (Internet Protocol DataCasting). With IP Datacast, content is delivered in the form of data packets using the same distribution technique as used for delivering digital content on the Internet. The use of Internet Protocol to carry its data, in so-called IP packets, allows DVB-H to rely upon standard components and protocols for content manipulation, storage and transmission. In addition to video and audio stream broadcasting, IP Datacast over DVB-H system can be used also for file delivery.
   e. Overview of the system

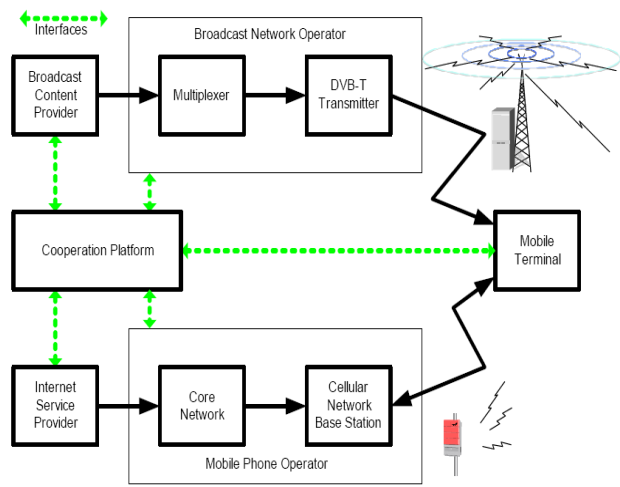

**Figure 5 System architecture for collaboration between mobile and broadcast operators**

   f. Optional characteristics. Broadcast services can be delivered by DVB-H without the need for an interaction channel, or in the configuration shown, an interaction channel can easily be provided by the use of a cellular network such as the GSM network. Methods of providing payment for services can be built upon a proprietary encryption and payment solution or in conjunction with the telecoms network's inherent service statistics collection and billing functions. The DVB Project has been elaborating these options in the Convergent Mobile and Broadcast Services (CMBS) group.

6. **Network architecture**

While the DVB-T network is intended primarily for roof top antenna reception, a DVB-H network will be designed for portable reception available even inside buildings. Hence it will need a much higher signal power density. In order to reach the higher power density needed for mobile coverage levels, several network architectures can be used depending on available frequencies, allowed maximum transmitter powers and antenna heights. The following network scenarios are possible:
- a. Existing DVB-T network with indoor coverage and DVB-H within the same multiplex,
- b. DVB-T and DVB-H using hierarchical transmission in the same radio frequency channel with DVB-H on the high-priority stream, or
- c. A DVB-H only network (which can then make use of the optional 4K mode if needed).
- d. DVB-H can use both Single Frequency Networks (SFN) and Multiple Frequency Networks (MFN) topologies.

## 7. One main transmitter and several repeaters

The simplest network architecture is one that uses a main transmitter with several repeater transmitters to boost the signal level at the edges of the cell. These repeaters may be necessary when it is not possible to have a high tower for the main transmitter or to fill-in shadows in the reception pattern.

A repeater is a special high gain antenna amplifier that takes the input signal via a receiving antenna, amplifies it and connects the signal to a transmitter antenna. This kind of network topology (essentially circular in shape) may not be very practical and experience shows that several transmitters may be required, each extended by some repeaters, to encompass the entire coverage area required.

## 8. Single Frequency Network (SFN)

An efficient network for DVB-H reception can be built by using several transmitters on the same frequency. A large area of up to 60 kilometers can be covered without needing high transmitter towers. The identical signals are transmitted from several sites and the system ehavior is similar to that of a distributed transmitter. The DVB-H main transmitters must be accurately synchronized, most easily with time signals received from GPS satellites. Repeaters can be used to improve coverage on critical areas where indoor or car reception performance has been found to be insufficient. This kind of network structure is sometimes known as a Dense SFN network.

## 9. Nation-wide coverage[4]

When nationwide coverage is required, over distances of hundreds of kilometers, several radio frequency channels will be needed. The availability of channels differs very much from one country to another. In theory, three channels should be sufficient to provide continuous coverage with any area. However, practical network planning shows that 5-6 channels are actually needed. By using different channels in neighbouring areas gives the possibility also to run local content in each area. This may be important with DVB-H where local content is expected to have an important role.

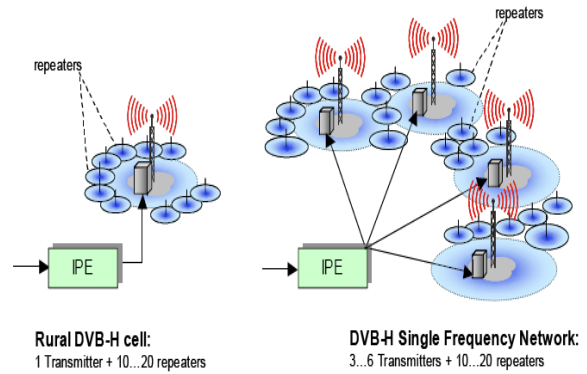

**Figure 6 Possible network topology solutions for DVB-H**

## 10. Conclusions

Traditional broadcasting is undergoing a process of change as a consequence of the move towards an all-digital broadcasting environment. New technology, such as streaming technology and personal video recorders (PVRs) can complement traditional broadcasting. New players are entering the market. More programs, competition and new distribution platforms means that the television viewing experience will change. For television providers, the arrival of IPTV and the enhanced offering of cable and satellite providers has increased the competition. However the terrestrial platform benefits from a unique competitive advantage – that of wireless mobility.

New technologies, such as UMTS, are enabling mobile telecom operators to provide television like services to their subscribers and enter the television broadcast market. Already, this has led to nascent cooperation between broadcasters and mobile operators. However, because UMTS networks cannot provide television-like services to a large population at a reasonable cost, these services will likely become available via a broadcast network. In order to retain a role in the provision of television services to handheld devices, broadcasters will need to stake their claim quickly or risk the involvement of new players in the market.

Because the ideal spectrum for DVB-H services is assigned to broadcasting, using a DVB-H network enables broadcasters to retain a leading role and leverage their strengths in the provision of content. But broadcasting television services to a handheld using the DVB-H standard will require compromises among the players. The technology to provide handheld television services exists. Consumer demand for such services is expected to grow, and it may be possible to commercially launch such services as early as 2006. However, key regulatory and business issues will need to be resolved. Broadcasters and other members of the value chain should use this time to consider how handheld television services such as DVB-H may be integrated into their strategy.